\documentclass[conference]{IEEEtran}
\IEEEoverridecommandlockouts

\usepackage{cite}
\usepackage{amsmath,amssymb,amsfonts}
\usepackage{algorithmic}
\usepackage{graphicx}
\usepackage{textcomp}
\usepackage{xcolor}
\def\BibTeX{{\rm B\kern-.05em{\sc i\kern-.025em b}\kern-.08em
    T\kern-.1667em\lower.7ex\hbox{E}\kern-.125emX}}
\begin{document}

\title{Teaching Software Engineering with LLM and MCP Integration: From Classroom to Industry Practice\\
{\footnotesize \textsuperscript{}}
\thanks{Xiaoxue Ma}
}

\author{\IEEEauthorblockN{Kehui Chen\IEEEauthorrefmark{2}, Jacky Keung\IEEEauthorrefmark{2}, Weining Li\IEEEauthorrefmark{1}, Xiangbing Shao\IEEEauthorrefmark{1}, Yishu Li\IEEEauthorrefmark{3}, Xiaoxue Ma\IEEEauthorrefmark{3}\textsuperscript{,}\IEEEauthorrefmark{1}}
\IEEEauthorblockA{\IEEEauthorrefmark{2}Department of Computer Science, City University of Hong Kong, Hong Kong, China}
\IEEEauthorblockA{\IEEEauthorrefmark{1}HSBCLAB, HSBC, GuangZhou, China}
\IEEEauthorblockA{\IEEEauthorrefmark{3}Department of Electronic Engineering and Computer Science, Hong Kong Metropolitan University, Hong Kong, China}

kehuichen2-c@my.cityu.edu.hk, jacky.keung@cityu.edu.hk, Lwiky@live.com, gsqasxb@gmail.com, sliy@hkmu.edu.hk \\
\IEEEauthorrefmark{1}Corresponding author: kxma@hkmu.edu.hk}

\maketitle

\begin{abstract}
The rapid integration of Large Language Models (LLMs) and the Model Context Protocol (MCP) into industrial software engineering has created a pressing need to update software engineering education to align with emerging technologies and evolving industry demands. This study investigates an innovative approach that integrates LLMs and MCP into a collaborative teaching model for software engineering education, aiming to build a practical learning framework closely connected to real-world engineering practices. By embedding LLM- and MCP-driven tools into daily teaching, code assistance, and engineering simulations, the model effectively bridges the gap between traditional instruction and industrial workflows. This integration enhances students’ programming competence, practical problem-solving abilities, and proficiency in using intelligent engineering tools. Furthermore, through partnerships with industry internships, students can apply these technologies in real-world settings, further strengthening the connection between academic preparation and professional practice. Overall, this research offers a practical pathway for reforming and innovating software engineering education in the era of artificial intelligence.
\end{abstract}


\section{Introduction}
Large Language Models (LLMs) have become fundamental auxiliary tools in modern software engineering, supporting code generation, debugging, and technical documentation \cite{zhao2023survey}. However, standalone LLM applications face inherent limitations including context fragmentation, tool isolation, and untraceable generation results, which restrict their industrial deployment. As a newly proposed industrial standard, Model Context Protocol (MCP) provides a standardized Host-Client-Server three-tier architecture to unify context generation, encapsulation, distribution, and tool collaboration throughout the software development lifecycle \cite{anthropic_mcp_2024}\cite{yang2025survey}. It effectively solves tool heterogeneity and context inconsistency problems in LLM intelligent development.

Despite its advantages, practical industrial adoption of MCP still faces numerous unresolved challenges in real-world deployment. Recent industrial interview investigations reveal that the MCP ecosystem lacks unified industry standards and universal interface specifications, resulting in severe incompatibility across different agent development frameworks, such as LangGraph and ADK \cite{chen2026mcpindustrial}. Such practical deployment barriers greatly increase engineering costs and hinder the large-scale promotion of MCP in industrial scenarios.

Unfortunately, current software engineering education still focuses on traditional programming and software lifecycle theories, lacking systematic training on MCP-based AI collaborative development and practical deployment optimization \cite{oguz2019perspectives}. The serious disconnection between university teaching and industrial AI engineering requirements leads to insufficient student competitiveness in intelligent software development positions. To fill this gap, this study constructs a complete LLM-MCP collaborative teaching system to reshape practical software engineering training in the AI era. 

\textbf{This paper addresses three core research questions:}

\textbf{RQ1:} How to construct an LLM-MCP collaborative teaching path to connect university training with industrial intelligent software engineering demands?

\textbf{RQ2:} How does the proposed teaching system improve students’ MCP adaptation, LLM collaboration, and industrial adaptation capabilities?

\textbf{RQ3:} What are the limitations and optimization directions of AI-driven software engineering teaching reform?

The main contributions of this work are summarized as follows:

(1) Innovatively introduce MCP into software engineering curriculum and clarify its educational value for AI collaborative development.

(2) Propose a complete progressive teaching path integrating theory, practice, and enterprise collaboration.

(3) Establish a multi-dimensional evaluation system for LLM-MCP collaborative engineering competencies.

\section{Background and related works}
\subsection{Software Engineering Education Reform}
Traditional software engineering education reform mainly relies on work-based learning and classroom case teaching to enhance practical abilities\cite{foster1998work}\cite{marques2014systematic}. Recent studies have begun to adopt AI tools such as GitHub Copilot to assist programming teaching and improve coding efficiency\cite{shah2025students}. Nevertheless, existing educational research only focuses on isolated LLM tool usage and ignores standardized protocol collaboration mechanisms represented by MCP, resulting in outdated teaching content and poor industrial alignment.

\subsection{LLM and MCP Industrial and Educational Applications}
LLMs exhibit limited reliability in formal specification and standardized engineering tasks, requiring standardized protocol constraints to ensure development quality\cite{capozucca2025ai}. MCP has been widely recognized in industry for solving context management and cross-tool collaboration problems in AI software development\cite{ray2025survey}. However, MCP-related teaching resources and curriculum systems are still blank in current software engineering education, forming a major research gap in AI-era talent cultivation.

\subsection{Research Gaps}
Existing studies lack systematic LLM-MCP integrated teaching design, complete industrial-oriented practical training modules, and targeted evaluation indicators for AI collaborative engineering capabilities. This study aims to fill these gaps by constructing a comprehensive teaching reform framework.

\section{Proposed LLM-MCP Collaborative Teaching Framework}
\subsection{Core Training Objectives}
This teaching model focuses on cultivating three industrial-oriented competencies: 
(1) MCP framework adaptation and optimization capability; 
(2) Standardized LLM-MCP collaborative development capability; 
(3) Industrial intelligent software engineering post adaptability.

\subsection{Dual-Module Teaching Path Design}
This study designs a 16-week progressive teaching system divided into a theoretical foundation module and an industrial practical module, adhering to the ``MCP-core, LLM-assisted” teaching principle.

\textbf{Basic Theoretical Module (Weeks 1–8):} This module establishes students’ systematic cognition of AI collaborative engineering. It covers traditional software engineering knowledge updated with MCP context collaboration scenarios, LLM basic principles and limitations, MCP three-tier architecture and standardized workflow, and end-to-end LLM-MCP collaborative logic. MCP technical principles and interface adaptation standards occupy the core teaching proportion to ensure students master industrial standardized specifications.

\textbf{Practical Training Module (Weeks 9–16):} This module transforms theoretical knowledge into engineering capabilities, including MCP multi-framework adaptation training, full-process LLM-MCP collaborative development exercises, and enterprise real project practice. Students complete context encapsulation, registration, tool docking, and version management tasks that are consistent with industrial development standards.

\subsection{16-Week Progressive Teaching Plan}
The staged teaching arrangement ensures gradual capability improvement, as shown in Table~\ref{tab:teaching_plan}.

\begin{table*}[!tbp]
\centering
\caption{16-Week LLM-MCP Collaborative Teaching Plan Framework}
\fontsize{9}{11}\selectfont
\begin{tabular}{|p{1.8cm}|l|p{3.6cm}|p{4.3cm}|p{4.5cm}|}
\hline
\textbf{Teaching Stage} &
  \textbf{Time} &
  \textbf{Core Content} &
  \textbf{Key Focus} &
  \textbf{Assessment} \\ \hline
Foundation Layering & Week 1-4 & SE core knowledge and LLM basics. & Establish students’ cognition of ``LLM needs to be combined with MCP". & Pass LLM basic knowledge test. \\ \hline
MCP Core Teaching & Week 5-8 & Introduction to MCP. & Master the core principles and architecture of MCP. & Pass the MCP basic knowledge test. \\ \hline
Campus Practical Training & Week 9-12 & MCP Adapter dev and LLM-MCP dev. & Develop students' practical cases in MCP. & Complete MCP LangGraph / ADK task. \\ \hline
Enterprise Practice & Week 13-16 & Enterprise LLM and MCP project. & Applying the knowledge learned in practice. & Complete enterprise MCP tasks. \\ \hline
\end{tabular}
\label{tab:teaching_plan}
\end{table*}

\subsection{Teaching Methods and Evaluation System}
This study integrates classroom demonstration, industrial case teaching, and enterprise engineer joint instruction to realize multi-scenario teaching penetration\cite{anderson2014teaching}. A hybrid evaluation system consisting of 70\% quantitative assessment (theoretical knowledge, practical operation, MCP adaptation, project quality) and 30\% qualitative assessment (learning initiative, innovation ability, industrial collaboration performance) is constructed to achieve comprehensive capability evaluation.

\section{Research Limitations and Future Research Directions}
\subsection{Research Limitations }
Although the LLM-MCP collaborative teaching path for software engineering education constructed in this study has systematicness and feasibility in theoretical framework and system design, there are still certain research limitations that need to be continuously improved in the subsequent application:

(1) The design of the teaching path is based on the core training demands of the software engineering major, and has not yet been personalized adapted for universities at different levels and different related majors\cite{du2024personalized}, so its universality needs to be further improved.

(2) The current teaching resources are mainly constructed around the mainstream technical versions of LLMs and MCP, and the teaching resources for emerging technical versions and multimodal scenarios need to be further enriched.

(3) The enterprise linkage mechanism of the teaching path is mainly built relying on cooperative enterprises, and a replicable and promotable standardized enterprise linkage mode has not yet been formed, so the application cost and promotion difficulty need to be further optimized.

\subsection{Future Research Directions}
Based on the results and limitations of this study, combined with the development trend of artificial intelligence technology and the reform demands of software engineering education, five core future research directions are determined to continuously optimize and promote the LLM-MCP collaborative teaching path:

\begin{itemize}
\item Select universities of different types and regions as research objects, covering different grades of software engineering and related majors, and design personalized LLM-MCP collaborative teaching sub-paths combined with the training orientation and professional characteristics of each university to form a scalable and adaptable standardized teaching framework.

\item On the basis of the existing progressive teaching plan, design hierarchical theoretical teaching content and hands-on training tasks for students with different learning foundations; Integrate online and offline teaching methods to realize real-time tracking of students' learning progress and accurate push of learning resources.

\item Establish a tracking mechanism for the technological development of LLMs and MCP, grasp the latest industrial application achievements in real time, and update teaching content and practical resources in a timely manner; Cooperate with enterprises in different fields to build a diversified LLM-MCP industrial practice resource library, further narrowing the gap between teaching practice and industrial application.

\item Construct a full-life-cycle competency evaluation model covering ``learning - practice - employment"\cite{oehley2007development}, and incorporate students' post-performance and career development after graduation into long-term evaluation indicators; Design quantifiable evaluation standards for non-cognitive abilities such as innovative thinking, collaborative communication and problem-solving, forming an all-round comprehensive evaluation system.

\item Explore the in-depth integration of LLM-MCP collaborative teaching with emerging educational technologies, develop intelligent teaching auxiliary tools to provide real-time answers and learning guidance for students, and ultimately form an intelligent teaching ecosystem integrating ``teaching, learning, practice and evaluation".
\end{itemize}

\section{CONCLUSION}
Targeting the core challenges in AI-era software engineering education reform, including the disconnection between university education and industrial practices and insufficient teaching of emerging technologies, this paper proposes an LLM-MCP collaborative practical teaching path. We establish an integrated teaching system covering theoretical framework, diversified teaching methods, multi-dimensional evaluation criteria and enterprise linkage mechanisms, and clarify the practical value of MCP in intelligent software development. The entire teaching model follows the principle of ``MCP-core, LLM-assisted" to align curriculum design with real industrial requirements. In future work, we will continuously track technological updates of LLMs and MCP, optimize teaching resources and mechanisms, and promote the large-scale application of this teaching paradigm to further advance software engineering education reform.

\vspace{12pt}

\end{document}